%% file: LCSS2020.tex
\documentclass[letterpaper, 10 pt, journal,twoside]{ieeetran}

\IEEEoverridecommandlockouts                              

\usepackage{graphics} 
\usepackage{epsfig} 

\makeatletter
\let\proof\@undefined                        
\let\endproof\@undefined                  
\makeatother

\usepackage{amsmath,amssymb,amsfonts,amsthm}
\usepackage{algorithmic}
\usepackage[linesnumbered,ruled,vlined,onelanguage]{algorithm2e}
\usepackage{array}
\usepackage{dsfont,color}
\usepackage{caption}
\usepackage{subcaption,balance}

\newtheorem{prop}{Proposition} 

\newtheorem{thm}{Theorem}

\newtheorem{defn}{Definition}

\newtheorem{example}{Example}
\newtheorem{problem}{Problem}

\newcommand{\hasnb}[1]{{\color{black} #1}}
\newcommand{\syong}[1]{{\color{black} #1}}
\newcommand{\qiang}[1]{{\color{black} #1}}

\title{\LARGE \bf
Equalized Recovery State Estimators for \hasnb{Linear} Systems with Delayed and Missing Observations
}

\author{Syed M. Hassaan, Qiang Shen and Sze Zheng Yong 
\thanks{S.M. Hassaan and S.Z. Yong are with the School for Engineering of Matter, Transport and Energy, Arizona State University, Tempe, AZ, USA. Q. Shen is with the School of Aeronautics and Astronautics, Shanghai Jiao Tong University, Shanghai, P.R. China (e-mail: \{shassaan,szyong\}@asu.edu, qiangshen@sjtu.edu.cn). This  work  was  supported  in  part  by  DARPA grant D18AP00073 and NSF grant CNS-1943545.}%
}

\begin{document}

\maketitle
\thispagestyle{empty}
\pagestyle{empty}

\input{my_abstract}

\vspace{-0.3cm}
\input{my_intro}

\input{my_prob}

\input{my_design}

\input{my_example}

\vspace{-0.1cm}
\section{Conclusions} 
\vspace{-0.05cm}
In this paper, we focused on the problem of synthesizing a dynamic state observer that has the ability to achieve equalized recovery when a discrete-time \hasnb{linear} time-varying system is subjected to delayed or missing data in a finite time horizon. 
{To achieve this, we constructed} a reduced event-based language capable of capturing {the} set of indistinguishable event {sequences} that different delay mode {sequences in the delayed data language correspond to, and}  
augmented {associated constraints to a novel} 
equalized recovery estimator constructed with time-varying intermediate levels {and the recovery level as a decision variable. When compared to existing designs, our proposed estimator can adapt the estimator gains at run time based on observed prefix of the language and can directly optimize the recovery and intermediate levels, leading to improved estimation performance.} 
\normalsize
\vspace{-0.1cm}
\bibliographystyle{IEEEtran}
\bibliography{biblio}

\normalsize
\vspace{-0.2cm}
\appendix \vspace{-0.1cm}
Matrices and vectors in Theorem \ref{th:delayed} are defined as follows:

\vspace{-0.35cm} \small{\begin{gather*} 
W \hspace{-0.08cm} = \hspace{-0.08cm}
\left[  \setlength\arraycolsep{1.1pt}
\begin{array}{cccc}
W_0   			&	\cdots	&   0 		\\[-5pt]
\vdots  &      \ddots 	&  \vdots 		\\
0		&   \cdots	  	&   W_{T-1}
\end{array}
\right], V \hspace{-0.08cm} = \hspace{-0.08cm}
\left[  \setlength\arraycolsep{1.1pt}
\begin{array}{cccc}
V_0   			&	\cdots	&   0 		\\[-5pt]
\vdots  &      \ddots 	&  \vdots 		\\
0		&   \cdots	  	&   V_{T-1}
\end{array}
\right],
\\[-5pt]
A \hspace{-0.1cm} =  \hspace{-0.12cm}
\left[  \setlength\arraycolsep{1.1pt}
\begin{array}{c}
I_n 	 	\\
A^1_0   	\\[-5pt]
\vdots  	\\
A^{T}_0 
\end{array}
\right]\hspace{-0.1cm}, C  \hspace{-0.1cm} =  \hspace{-0.12cm}
\left[  \setlength\arraycolsep{1.1pt}
\begin{array}{ccccc}
C_0   	&	0		&	\cdots	&   0 		& 0 		\\[-5pt]
0	 	&   C_1   	&   \ddots 	&   \vdots	& \vdots 	\\[-5pt]
\vdots  &   \ddots 	&   \ddots 	&   0 		& 0 		\\
0		&   \cdots	&   0   	&   C_{T-1} & 0
\end{array}
\right]\hspace{-0.1cm}, 
{H} \hspace{-0.1cm}=\hspace{-0.12cm}
\left[ \setlength\arraycolsep{1.2pt}
\begin{array}{ccccc}
0   	&	0		&	0		&   \cdots	&  0 		\\
A^1_1 	&   0   	&   0   	&   \cdots	&  0 		\\[-5pt]
A^2_1   &   A^2_2 	&   0   	&   \cdots	&  \vdots 	\\[-5pt]
\vdots  &   \vdots	&   \ddots 	&   \ddots	&  0   		\\
A^T_1 	&   A^T_2	&  A^T_3	&   \cdots  &  A^T_T
\end{array}
\right]\hspace{-0.1cm},\\[-5pt]
M^{\alpha}  \hspace{-0.08cm} =  \hspace{-0.08cm} \left[ \setlength\arraycolsep{1.1pt}
\begin{array}{cccc}
M_{(0,0)}^{\alpha}	&	0			&	\cdots	&	0		\\[-5pt]
M_{(1,0)}^{\alpha}	&	M_{(1,1)}^{\alpha}	&	\ddots	&	\vdots	\\[-5pt]
\vdots		&	\vdots		&	\ddots	&	0		\\
M_{(T-1,0)}^{\alpha}	&	M_{(T-1,1)}^{\alpha}	&	\cdots	&	M_{(T-1,T-1)}^{\alpha}
\end{array}
\right],
\Phi^{\alpha}  \hspace{-0.08cm} =  \hspace{-0.08cm}
\left[
\begin{array}{c}
I_n 	 	 \\
\Phi^{{\alpha},1}_0\\[-5pt]
\vdots  	 \\
\Phi^{{{\alpha},T}}_0 
\end{array}
\right], 
\\[-5pt]
L^{\alpha}  \hspace{-0.08cm}= \hspace{-0.08cm}
\left[ \setlength\arraycolsep{1.1pt}
\begin{array}{cccc}
L_0^{\alpha}   	&	0		&	\cdots	&   0 	 	\\[-5pt]
0	 	&   L_1^{\alpha}   	&   \ddots 	&   \vdots 	\\[-5pt]
\vdots  &   \ddots 	&   \ddots 	&   0 	 	\\
0		&   \cdots	&   0   	&   L_{T-1}^{\alpha}
\end{array}
\right], 
\Gamma^{\alpha} \hspace{-0.08cm}=\hspace{-0.08cm}
\left[ \setlength\arraycolsep{1.1pt}
\begin{array}{ccccc}
0   		&	0			&	0		&   \cdots	&  0 		\\
\Phi^{{\alpha},1}_1 	&   0   		&   0   	&   \cdots	&  0 		\\[-5pt]
\Phi^{{\alpha},2}_1   	&   \Phi^{{\alpha},2}_2 	&   0   	&   \cdots	&  \vdots 	\\[-5pt]
\vdots  	&   \vdots		&   \ddots 	&   \ddots	&  0   		\\
\Phi^{{\alpha},T}_1 	&   \Phi^{{\alpha},T}_2	&  \Phi^{{\alpha},T}_3	&   \cdots  &  \Phi^{{\alpha},T}_T
\end{array}
\right]\hspace{-0.1cm}, \end{gather*}} \normalsize \vspace*{-0.15cm}

\noindent for all $ \alpha \in \mathbb{N}_1^{| \mathcal{L}^{E^{\prime} }|}$, where $A^k_i = A_{k-1}A_{k-2}...A_i$,  $\Phi^{{\alpha},k}_i = \Phi_{k-1}^{\alpha} \Phi_{k-2}^{\alpha}...\Phi_i^{\alpha}$ and $\Phi_k^{\alpha} = A_k - L_k^{\alpha} C_k$.

\end{document}

%% file: my_abstract.tex

\begin{abstract} 
This paper presents a dynamic state observer design for discrete-time \hasnb{linear} time-varying  systems that robustly achieves \emph{equalized recovery} despite delayed or missing observations, where the set of all temporal patterns for the missing or delayed data is modeled by a finite-length language. By introducing a mapping of the language onto a reduced event-based language, {we design a state estimator that  adapts based on the history of available data at each step, and satisfies equalized recovery  for all patterns in the reduced language.} In contrast to existing equalized recovery estimators, the proposed design considers the equalized recovery level as a decision variable, which enables us to {directly} obtain the global minimum for the intermediate recovery level, resulting in improved estimation performance. Finally, we demonstrate the effectiveness of the proposed observer when compared to existing approaches using several illustrative examples.
\end{abstract}

\begin{IEEEkeywords}
Estimation; Delay systems; Observers for Linear systems
\end{IEEEkeywords}

%% file: my_intro.tex

\section{INTRODUCTION}

\IEEEPARstart{C}{yber-physical} systems (CPS) typically involve multiple sensors that send data packages to controllers through a shared communication channel, and controllers that compute and transmit control commands to actuators that are connected to the physical system. For the safe and efficient operation of these systems, state estimation plays an essential role. However, time delays and missing data are often inevitable due to sensor failures, package drops or adversaries.
Hence, there is a need for designing state estimators that are robust to these delays and missing data.

\emph{Literature review:} 
For the past few decades, active research development has been undertaken in the area of state estimation for systems that are susceptible to packet drops and delayed communication, as highlighted in \cite{zhang2001stability, jin2007estimation} as typical concerns in networked control systems. Significant amount of research has been done to design state estimators when only intermittent data is available \cite{sinopoli2004kalman, smith2003estimation, Battistelli2012}, and when observations are arriving as out-of-sequence measurements {\cite{zhang2012optimal,wang2017cooperative,matei2011optimal,zhang2011linear}}. The authors in \cite{zhang2012optimal} used complete in-sequence information approach to recompute all the estimations from the step when data did not arrive until the point when it finally arrived, while \cite{wang2017cooperative} proposed nonlinear filters utilizing a Bayesian filtering framework to correct the previous estimation as soon as the delayed observation arrives. On the other hand, an optimal state estimation approach was proposed for Markovian jump linear systems subject to delays in both the output 
and mode observations in \cite{matei2011optimal}.
However, these works mainly modeled the missing and delayed observations as stochastic variables with known probability distributions and focused on obtaining the best average/expected estimates as opposed to achieving \emph{best worst-case/robust estimation errors} considered in this paper.

Another relevant area that does consider the worst-case/robust estimation performance is the synthesis of set-valued estimators, which has seen some recent development, e.g., 
\cite{shamma1999set,milanese1991optimal,chen2005observer}. 
The authors in \cite{blanchini2012convex} introduced the property of equalized performance, which implies that the estimation error always remains equal/invariant. For systems with missing observations, \cite{Rutledge2018} and \cite{Hassaan2019ACC} modeled the feasible missing data patterns with a finite-length language and proposed finite-horizon affine estimators with an extended property called equalized recovery, which implies that within a finite time horizon, especially for times when  observations may go missing, the estimation error can have a more relaxed upper bound, but by the end of the horizon should return 
to the initial upper bound. In more recent work, \cite{Kwesi2019ACC,rutledge2020finite} developed a prefix-based method to predict the possible pattern of missing data to improve the estimation performance.  
However, this approach does not directly apply for delayed data patterns, and thus, our goal in this paper is to design equalized recovery estimators that can handle them. 

\emph{Contribution:} 
In this paper, we design a state observer that achieves equalized recovery when the system data is prone to misses and delays (including out-of-sequence observations). Instead of assuming probabilistic missing or delay events, we model them using 
a fixed-length language that represents the set of all possible temporal patterns of the missing or delayed data 
and further construct a reduced event-based language with unique event sequences. {In contrast to the worst-case language method in \cite{Rutledge2018, Hassaan2019ACC}, our proposed design monitors}
{
the history of available data 
{at run time and adapts the estimator} gain matrices. 
{Furthermore, we extend existing equalized recovery estimators to allow time-varying intermediate levels and}
consider the equalized recovery level as a decision variable, {enabling us to directly find the global minimum of the intermediate levels.} 
{These improvements} 
are shown in simulations to yield better estimation performance.} 

%% file: my_prob.tex

\vspace{-0.1cm}
\section{Problem Formulation}
\vspace{-0.0cm}

\subsection{System Dynamics and Delayed Data Language}
\vspace{-0.0cm}

\emph{System Dynamics:}
We consider a discrete-time \hasnb{linear} time-varying system subject to process noise and output noise. The model of the system dynamics is described as follows: 
\begin{align} \label{eq:system}
\begin{array}{rl}
		x_{k+1} &= A_kx_k + B_ku_k + W_kw_k,\\
		z_k &= C_k x_k + V_k v_k,  \\
		\hasnb{Y_k} &=\{ z_{k-\tau(i)} | i+\tau(i)=k, i \leq k\},
		\end{array}
\end{align}
where $x_k \in \mathbb{R}^n$ is the system state at time $k$ , $u_k \in \mathbb{R}^m$ is the input to the system, $w_k \in \mathbb{R}^n$ is the process noise, $v_k \in \mathbb{R}^p$ is the measurement noise, $z_k \in \mathbb{R}^p$ is the model output, $\hasnb{Y_k \subset \mathbb{R}^p}$ is \hasnb{\syong{the} set of all 
measurements\syong{/outputs} that are received at time step $k$ and}  
$\tau(i)\syong{\ge 0}$ is the \syong{unknown} time delay {of the data} at the time step $i$ that satisfies 
\syong{$\tau(i) \leq \bar{\tau}$, where} 
$\overline{\tau}$ is a known upper bound on the number of time steps that a packet \syong{can be} delayed by.  The discrete variable $\tau(i) = 0$ denotes that the 
measurement from time step $i$ is {received/}available, 
while $\tau(i) = \delta$ implies that the 
\syong{data from time step $i$} is delayed by $\delta$ steps.  We assume that $w_k$ and $v_k$ are bounded with $\|w_k\| \leq \eta_w$ and $\|v_k\|\leq \eta_v$ for \syong{each} 
$k$, 
\hasnb{where \qiang{$\|\cdot\|$} denotes the $\infty$-norm}. The system matrices $A_k$, $B_k$, $C_k$, $W_k$, $V_k$, \syong{$\eta_w$ and $\eta_v$} are all 
known.  Without loss of generality, we assume that the initial time is $k=0$.

\emph{Delayed Data Language:}
Given a fixed length $T$, we consider a delayed data model in which all delay patterns are restricted to a set expressed by fixed-length language specifications, e.g., `the $i$-th observation is delayed by at most $m$ time steps' or `at most $m$ available measurements in a fixed interval'. Formally, our delayed data model is a fixed-length language $\mathcal{L}$ \syong{of length $T$} that specifies the set of allowable delay mode 
sequences $\tau(0)\tau(1)\tau(2)\ldots\tau(T-1)$ \syong{with $\tau(i)\le \bar{\tau}, \forall i\in \mathbb{N}_0^{T-1}$}, where the $\alpha$-th possible sequence is 
called a word $\mathcal{W}_\alpha$ 
with $\alpha\in \mathbb{N}_1^{|\mathcal{L}|}$.  
Note that $i+\tau(i) \syong{\ge} T$ means that the $i$-th data 
is delayed beyond the horizon $T$, which is similar to the situation 
where that data is missing. In other words, the case considered in \cite{Rutledge2018, Hassaan2019ACC, Kwesi2019ACC} is a special case of the delayed data language in this paper. \vspace{-0.1cm}


\begin{example} \label{exm1}
Consider a system where the observation is delayed by at most $2$ time steps in a fixed interval of length 2. This means that $\overline{\tau} = 2$ and $T=2$, and hence we have $\tau(i) \in \{0,1,2 \}$ for all $i \in \mathbb{N}_0^{1}$. Therefore, the fixed-length language can be expressed as $\mathcal{L} = \{\mathcal{W}_1, \ldots, \mathcal{W}_9\} = \{00, 01, 02, 10, 11, 12, 20, 21, 22\}$.\vspace{-0.0cm}
\end{example}

\subsection{Equalized Recovery}

The focus of our paper is to design a bounded-error estimator, where the estimation error is guaranteed to return/recover to the same bound that it started with after a fixed number of time steps, as an extension of the notion of \emph{equalized performance} in \cite{blanchini2012convex}. In terms of time horizon $T$, we enforce that the estimation error bound at the end of the horizon is guaranteed to be less than or equal to the bound at the start. 
Formally,  we consider 
  \emph{equalized recovery}, 
{defined as follows,} 
{which is a slight modification of the definition in  \cite{Rutledge2018} to allow time-varying intermediate levels:} 

\begin{defn}[Equalized Recovery] 
\label{def:recovery}
	An estimator is said to achieve an equalized recovery level $\mu_1$ at time $0$ with recovery time $T$ and intermediate {levels} ${\mu_{2,k}}  \geq \mu_1$  if \syong{for any}  
	$\|\tilde{x}_{0}\| \leq \mu_1$, we must have $\|\tilde{x}_k\|\leq {\mu_{2,k}}$ for all $k \in [0, T]$ and $\|\tilde{x}_{T}\| \leq \mu_1$, where $\tilde{x}_k\triangleq x_k -\hat{x}_k$ is the estimation error and $\hat{x}_k$ is the state estimate at time $k$. 
\end{defn}

\subsection{Problem Statement} 
We aim to design a bounded-error estimator that satisfies \emph{equalized recovery}, 
which can be stated as follows:
{\begin{problem}[Estimator Design with Delayed Data]
\label{prob:EstimatorDesign}
	Given the system dynamics  (\ref{eq:system}), a delayed data model specified by a language ${\mathcal L}$ and a recovery time $T$ as a time horizon, design an optimal equalized recovery state estimator with estimate $\hat{x}_k$ and estimation error $\tilde{x}_k=x_k-\hat{x}_k, \forall k \in [0,T]$ that minimizes a cost {$J(\mu_1,\{\mu_{2,k}\}_{k=0}^T)$}
	subject to $\mu_{2,k} \geq\mu_1$, $\|\tilde{x}_k\| \leq \mu_{2,k}, \; \forall k \in [0,T]$ 
	 and $\|\tilde{x}_{T}\|\leq \mu_1$ \syong{for all $\|\tilde{x}_{0}\|\leq \mu_1$}.
\end{problem}}

{In contrast to \cite{Rutledge2018,Hassaan2019ACC,Kwesi2019ACC,rutledge2020finite}, the above problem formulation allows time-varying intermediate levels and more importantly, the equalized recovery level $\mu_1$ does not need to be specified \emph{a priori}. As a result, we can directly optimize over {$J(\mu_1,\{\mu_{2,k}\}_{k=0}^T)$}  and overcome the challenge with the formulation in \cite{Rutledge2018,Hassaan2019ACC,Kwesi2019ACC,rutledge2020finite} that the optimal cost is not monotonic in $\mu_1$. Furthermore, if the recovery time $T$ is not given and can be chosen, we can perform a line search over $T$ using the above formulation.} \syong{Moreover, simple modifications will allow us to consider affine dynamics, similar to \cite{Rutledge2018,Hassaan2019ACC,Kwesi2019ACC,rutledge2020finite}, and the case when the initial estimation error is greater than $\mu_1$ (cf. Section \ref{sec:imp}). }

%% file: my_design.tex

\section{Design Approach}
In this section, we propose an observer design approach to solve Problem \ref{prob:EstimatorDesign}, which involves constructing an event-based language $\mathcal{L}^E$ from the fixed-length delayed data language $\mathcal{L}$ and designing an estimator that adapts to the information from the {observed data pattern seen so far}.  

\input{set_my_tree}

\input{estimator_intro}

\input{my_theorem}

\input{my_robustification}

%% file: set_my_tree.tex
\vspace{-0.05cm}
\subsection{Event-Based Language}

Given a language set $\mathcal{L}$ of a system, containing all possible words for different allowable delay mode sequences, an event-based language $\mathcal{L}^E$ is constructed to capture the set of indistinguishable event {sequences} that {correspond to} the different delay mode {sequences in $\mathcal{L}$.} 
To build the {event-based language,} 
the following definitions are introduced first: 

\begin{defn}[Event] \label{defn:event}
An event $e_{i,j} = d_0d_1d_2 \ldots d_{i}$ at time step $i \in \mathbb{N}_0^{T-1}$ is a finite sequence of binary variables $d_l \in \{0,1\}$ for all $l \in \mathbb{N}_0^{i}$, where $j \in \mathbb{N}_0^{2^{i+1}-1}$ is an index denoting the $j$-th potential event at time step $i$. The binary variable $d_l = 1$ denotes that the data of time step $l$ is available at current time step $i$ (i.e., all received data up until the current step i), while $d_l = 0$ signifies that the data of time step $l$ is not available at current time step $i$. Moreover, an event can be {defined using} 
$e_{i,j} =\texttt{{\textup{binary}}}(j,i+1)$ at time step $i$, where the function $\texttt{{\textup{binary}}}$ returns a binary representation of the number $j \in \mathbb{N}_0^{2^{i+1}-1}$ with $i+1$ digits.\vspace{-0.1cm}
\end{defn}

\begin{defn}[Event Set] \label{defn:eventset}
An event set $e_{i} = \{ e_{i,j} \}_{j=0}^{2^{i+1}-1}$ is a set of all potential events at time step $i \in \mathbb{N}_0^{T-1}$. \vspace{-0.1cm}
\end{defn}

Intuitively, an event at time step $i$ represents the information 
that is available up until time 
$i$.
Since any data from previous or current steps 
only has two possibilities, i.e., measured or not measured at the current time $i \in  \mathbb{N}_0^{T-1}$, there is a total of $2^{i+1}$ different cases. Thus, the index $j$ of $e_{i,j}$ varies from $0$ to $2^{i+1}-1$, as exemplified in the following. 

\begin{example} \label{exm2}
Consider a system with a 
horizon $T= 2$. Based on above definitions of event and event set, we have $e_0 = \{ e_{0,j} \}_{j=0}^{2^1-1}= \{e_{0,0}, e_{0,1}\} = \{0, 1\}$ for time step $i=0$, $e_1 = \{ e_{1,j} \}_{j=0}^{2^2-1} = \{e_{1,0}, e_{1,1}, e_{1,2}, e_{1,3}\} = \{00, 01, 10, 11\}$ for time step $i=1$.
For instance, the event $e_{0,0} = 0$ means that the data of time $0$ is not available at the time $0$, and the event $e_{1,1} = 01$ means that at the time $1$, the data of time $0$ is not available but the data of time $1$ is available.\vspace{-0.1cm} 
\end{example}

\begin{defn}[Event Sequence] \label{defn:eventseq}
An event sequence $\mathcal{E}_ \alpha=e_{0,j_0}e_{1,j_1}e_{2,j_2} \ldots e_{T-1,j_{T-1}}$ is a sequence of  events {corresponding to} a word $\mathcal{W}_ \alpha = \{\tau(i)\}_{i=0}^{T-1}$ from the fixed-length language $\mathcal{L}$, where the subscripts $j_i$ for all $i \in \mathbb{N}_0^{T-1}$ are determined by the word $\mathcal{W}_ \alpha$.
\end{defn}\vspace{-0.1cm}

{In other words,} an event sequence represents the available information 
at each 
step.
For each 
delay mode sequence 
in a language $\mathcal{L}$, we can find its corresponding event sequence, and thus, 
the  language 
$\mathcal{L} = \{\mathcal{W}_j\}_{j=1}^{| \mathcal{L}|}$ containing all allowable delay mode sequences 
can be mapped onto an event-based language $\mathcal{L}^{E} = \{\mathcal{E}_\alpha\}_{\alpha=1}^{| \mathcal{L}|}$ containing all potential event sequences. Specifically, for a word $\mathcal{W}_\alpha = \tau(0)\tau(1)\tau(2)\ldots\tau(T-1)$, the subscript $j_k$, $k\in\mathbb{N}^{T-1}_0$, in the corresponding event sequence $\mathcal{E}_ \alpha=e_{0,j_0} e_{1,j_1}$ $e_{2,j_2} \ldots e_{T-1,j_{T-1}}$ (cf. Definition \ref{defn:eventseq}) can be constructed 
as
\begin{align} \label{eq:jk}
j_k = \textstyle\sum_{\ell = 0}^{k} 2^{\ell} \mathds{1}_{\tau(k-\ell) \le \ell}, \ \forall k\in\mathbb{N}^{T-1}_0,
\end{align}
where $\mathds{1}_{\tau(k-\ell)}$ denotes an indicator defined as \vspace{-0.1cm}
\begin{align}
\mathds{1}_{\tau(k-\ell) \le \ell} = \begin{cases} 1,\quad \tau(k-\ell) \le \ell, \\ 0, \quad \tau(k-\ell) > \ell.\end{cases}
\end{align}\vspace*{-0.2cm}

Note that the {resulting} event-based language $\mathcal{L}^{E} = \{\mathcal{E}_\alpha\}_{\alpha=1}^{| \mathcal{L}|}$ could have repeated event sequences {(i.e., the mapping is surjective)}. {Thus, we will eliminate} 
repeated event sequences in $\mathcal{L}^{E}$ {to obtain a reduced \emph{event-based language}} 
$\mathcal{L}^{E^{\prime}} =  \{\mathcal{E}_\alpha^\prime \}_{\alpha=1}^{| {L^{E^\prime}}|} \subseteq \mathcal{L}^{E}$ with unique event sequences $\mathcal{E}_\alpha^\prime$ for $\alpha \in \mathbb{N}_1^{| {L^{E^\prime}}|}$.
The next example demonstrates how to {map/}transform the fixed-length language {$\mathcal{L}$} in the Example \ref{exm1} to a reduced event-based language $\mathcal{L}^{E'}$. \vspace{-0.05cm}

\begin{example} \label{exm3}
Consider the delayed data language in Example \ref{exm1}. The word $\mathcal{W}_2 =01$ denoting that the data of time 0 has no delay while the data of time 1 is delayed by 1 time step, can be represented by the event trajectory $\mathcal{E}_2 =e_{0,1}e_{1,2}$ {using \eqref{eq:jk}}, where $e_{0,1} = 1$ means that the data of time $0$ is available at the time $0$, $e_{1,2} = 10$ means that the data of time $0$ is also available at the time $1$ (since the data of time 0 is previously received at the time $0$) and the data of time $1$ is not available at the time $1$ (since the data of time 1 is delayed by 1 time step). {Using this procedure, we can transform} 
all words in the language $\mathcal{L} = \{\mathcal{W}_1, \ldots, \mathcal{W}_9\} = \{00, 01, 02, 10, 11, 12, 20, 21, 22\}$ {to an event-based language} 
$\mathcal{L}^{E} = \{\mathcal{E}_1, \ldots, \mathcal{E}_9\} = \{e_{0,1}e_{1,3}, e_{0,1}e_{1,2}, e_{0,1}e_{1,2},$ $e_{0,0}e_{1,3}, e_{0,0}e_{1,2}, e_{0,0}e_{1,2}, e_{0,0}e_{1,1}, e_{0,0}e_{1,0}, e_{0,0}e_{1,0}\}$. 
Then, we can 
eliminate 
repeated event sequences in $\mathcal{L}^{E}$ and obtain a reduced event-based language $\mathcal{L}^{E'} = \{\mathcal{E}_1^{\prime}, \ldots, \mathcal{E}_6^{\prime}\} =\{e_{0,1}e_{1,3}, e_{0,1}e_{1,2},  e_{0,0}e_{1,3},  e_{0,0}e_{1,2}, e_{0,0}e_{1,1},$ $  e_{0,0}e_{1,0}\}$.\vspace{-0.1cm} 
\end{example}

%% file: estimator_intro.tex

\vspace{-0.15cm} 
\subsection{{Equalized Recovery} State Estimator {Design}}\label{sec:EstimatorIntro}
\vspace{-0.05cm} 

For the estimator design, we will make use of the following definitions and notation, inspired by \cite{Kwesi2019ACC}:\vspace{-0.125cm} 


\begin{defn}[\hasnb{Principal} Block Minor]
The $i$-th leading principal block minor of a 
matrix $M \in \mathbb{R}^{an \times bp}$ 
is the $n \times p$ block matrix,
$\mathcal{BM}_i(M) = M_{1:i n,1:i p}$,
for all $i \in [1, \min(a,b)]$. \vspace{-0.1cm}
\end{defn}

\begin{defn}[Prefix of an Event Sequence]
For an event sequence $\mathcal{E}_{\alpha}^{\prime} \in \mathcal{L}^{E^{\prime}}$ and $i \le | \mathcal{E}_{\alpha}^{\prime}|$, the length $i$ prefix of $\mathcal{E}_{\alpha}^{\prime}$ is defined as $\mathcal{E}_{\alpha}^{\prime, [1:i]} = e_{0,j_0}e_{1,j_1}e_{2,j_2} \ldots e_{i-1,j_{i-1}}$, where $|\mathcal{E}_{\alpha}^{\prime}|$ denotes the number of events in $\mathcal{E}_{\alpha}^{\prime}$. The set of all non-empty prefixes of $\mathcal{E}_{\alpha}^{\prime}$ is denoted as $Pref( \mathcal{E}_{\alpha}^{\prime})$.\vspace{-0.1cm}
\end{defn}

\begin{example}
Consider event $\mathcal{E}_1^{\prime}$ of the reduced event-based language $\mathcal{L}^{E^{\prime}}$ in Example \ref{exm3}. As $\mathcal{E}_1^{\prime} = e_{0,1} e_{1,3}$, we have $|\mathcal{E}_1^{\prime} | =2$ and $i = \{ 1,2\}$. The length 1 prefix of $\mathcal{E}_1^{\prime}$ is $e_{0,1}$, while its length 2 prefix is $e_{0,1}e_{1,3}$. Thus, the set of non-empty prefixes of $\mathcal{E}_1^{\prime}$ is $Pref( \mathcal{E}_{1}^{\prime}) = \{e_{0,1}, e_{0,1}e_{1,3}\}$. \vspace{-0.1cm}
\end{example}

{Next,} to solve 
Problem \ref{prob:EstimatorDesign}, we consider a finite horizon dynamic state estimator, inspired by \cite{Hassaan2019ACC}, 
with augmented states $\bar{x}_k \triangleq \begin{bmatrix} \hat{x}_k^\top & s_k^\top\end{bmatrix}^\top$, where $\hat{x}_k\in\mathbb{R}^n$ is the estimate of the system state and $s_k \in \mathbb{R}^n$ an auxiliary state which estimates $\tilde{x}_k{=x_k-\hat{x}_k}$. The estimator design is as follows:
\begin{align} \label{eq:estimator}
\begin{array}{rl}
\hat{x}_{k+1} &= A_k\hat{x}_k + B_ku_k - u_{e,k}, \\
s_{k+1} &= A_ks_k + u_{e,k} + {L_{k}^{\mathcal{E}_{\alpha}^{\prime}}} \syong{\tilde{z}_{k},}
 \end{array}
\end{align}
\syong{with} \vspace*{-0.5cm}
\begin{gather*}
\syong{\hspace*{0.95cm}\tilde{z}_{k} \hspace{-0.05cm}=\hspace{-0.05cm} \begin{cases}
\tilde{y}_{k} \hspace{-0.05cm}-\hspace{-0.05cm} C_k s_k \hspace{-0.05cm}=\hspace{-0.05cm} {z}_{k} \hspace{-0.05cm}-\hspace{-0.05cm} C_k (\hat{x}_k\hspace{-0.075cm}+\hspace{-0.075cm}s_k), & \hspace{-0.12cm} \text{if } z_{k} \hspace{-0.075cm}\in\hspace{-0.075cm} \bigcup_{j=0}^k Y_{j},\\
0, & \hspace{-0.12cm} \text{otherwise},\end{cases}} 
\end{gather*} \vspace*{-0.2cm}

\noindent where \syong{$\tilde{y}_k \triangleq z_{k}-C_k \hat{x}_k$},  ${L_{k}^{\mathcal{E}_{\alpha}^{\prime}}} \in \mathbb{R}^{n \times p}$ is the Luenberger gain {at step $k$ as 
a function of the observed prefix $\mathcal{E}_{\alpha}^{\prime}$} and $u_{e,k} \in \mathbb{R}^n$ is the \emph{causal} output error injection term given by: 
\begin{align} \label{eq:affine_inp}
u_{e,k} = {\nu_{k}^{\mathcal{E}_{\alpha}^{\prime}}} + \textstyle\sum_{i=0}^{k} {M_{(k,i)}^{\mathcal{E}_{\alpha}^{\prime}}} \syong{\tilde{z}_i}, 
\end{align} 
\noindent where 
${M_{(k,i)}^{\mathcal{E}_{\alpha}^{\prime}}} \in \mathbb{R}^{n \times p}$ and ${\nu_{k}^{\mathcal{E}_{\alpha}^{\prime}}} \in \mathbb{R}^n$ are gain matrices {at time $k$ as 
a function of the observed prefix $\mathcal{E}_{\alpha}^{\prime}$}, which will be designed to satisfy 
the objectives of Problem \ref{prob:EstimatorDesign}. 

In \cite{Hassaan2019ACC}, the estimator design was formulated with essentially one worst-case word in the worst-case language $\mathcal{L}^*$, which was obtained by combining all the words in the given language, resulting in a triplet of stacked $(M,L,\nu)$ matrices for the whole time horizon $T$ which would satisfy the conditions in Problem \ref{prob:EstimatorDesign} for both the worst-case word in $\mathcal{L}^*$ as well as individual words in $\mathcal{L}$. 
Since the worst-case language $\mathcal{L}^*$ is used for achieving the equalized recovery, the achievable performance level is conservative. 
On the other hand, solving Problem \ref{prob:EstimatorDesign} for multiple triplets of $(M^{\alpha},L^{\alpha},\nu^{\alpha})$ for each word $\mathcal{W}_{\alpha}$ in $\mathcal{L}$ may result in implementation conflicts due to causality. 
This limitation was discussed in detail in \cite{Kwesi2019ACC}.

To remedy this, 
we need to design the individual $(M^{\alpha},L^{\alpha},\nu^{\alpha})$ for each word  $\mathcal{W}_{\alpha}$ of $\mathcal{L}$ such that if two different words are not distinguishable until 
time $\bar{k}$, then 
$(M^{\alpha}_{(k)},L^{\alpha}_{(k)},\nu^{\alpha}_{(k)})$ for both words should be constrained to be the same for all $k \in \mathbb{N}_0^{\bar{k}-1}$, where $M^{\alpha}_{(k)}$ denotes the $k$-th row of $M^{\alpha}$.  
Instead of associating a triplet $(M^{\alpha},L^{\alpha},\nu^{\alpha})$ to each word $\mathcal{W}_{\alpha}$ in the language $\mathcal{L}$, we only consider triplets $(M^{\alpha},L^{\alpha},\nu^{\alpha})$ for each event sequence $\mathcal{E}_{\alpha}^{\prime}$ of the reduced event-based language $\mathcal{L}^{E^\prime}$. Since all event sequences in $\mathcal{L}^{E^\prime}$ are {not repeated and $\mathcal{L}^{E^\prime} \subseteq \mathcal{L}^{E}$}, 
we can reduce the number of triplets needed and thus the size of the optimization problem.

Using the above, 
we impose the following constraint due to indistinguishability of unique event trajectories in $\mathcal{L}^{E^{\prime}}$:

\small\vspace{-0.3cm}
\begin{equation} \label{eq:C} \hspace{0.1cm}
\mathcal{C}(\mathcal{L}^{E^{\prime}}) \hspace{-0.08cm}  = \hspace{-0.08cm} \left\{ \begin{matrix} \begin{matrix} \{(M^{\alpha},  L^{\alpha}, \\ {\nu}^{\alpha})\}^{\vert\mathcal{L}^{E^{\prime}}\vert}_{\alpha=1} \end{matrix} \left \vert 
\begin{matrix}  (e\in Pref(\mathcal{E}_{\alpha}^{\prime}) \wedge e\in Pref(\mathcal{E}_{\beta}^{\prime} ) ) \\\Longrightarrow \quad
\forall \mathcal{E}_{\alpha}^{\prime}, \mathcal{E}_{\beta}^{\prime} \in \mathcal{L}^{E^{\prime}} : \\
(\mathcal{BM}_{\vert e\vert}(M^{\alpha})=\mathcal{BM}_{\vert e \vert}(M^{\beta}))\wedge\\ 
(\mathcal{BM}_{\vert e\vert}(L^{\alpha})=\mathcal{BM}_{\vert e \vert}(L^{\beta}))\wedge\\ 
((\nu^{\alpha})_{(1:\vert e \vert n)}=(\nu^{\beta})_{(1:\vert e\vert n)})
\end{matrix}  \right.
\end{matrix}\right\}\hspace{-0.05cm}. \hspace{-0.2cm}
\end{equation} \normalsize

{Intuitively, if any pair of event sequences share the same prefix of a particular length, then they are indistinguishable at the corresponding time step based on the received information. Since they are indistinguishable (and future information is inaccessible in a causal system), their associated submatrices and subvectors need to be constrained to be the same to avoid conflicts during implementation.} Note that while the prefix notation is similar to \cite{Kwesi2019ACC}, our estimator uses a different state estimator structure that enables us to consider more general data patterns, including delayed data patterns.

Moreover, {for each event sequence $\mathcal{E}_{\alpha}^{\prime} \in \mathcal{L}^{E^{\prime}}$,} due to delayed data and causality, all the entries in $M^\alpha$ and $L^\alpha$ corresponding to no available data 
should also be set to zero. To construct this constraint on $M^\alpha$ and $L^\alpha$, we first define an event matrix associated with the event sequence $\mathcal{E}_{\alpha}^{\prime} \in \mathcal{L}^{E^{\prime}}$: 

\small \vspace{-0.3cm}\begin{align*}
{E_{\alpha} = \begin{bmatrix} 
e_{0, j_0}^{(0)} 		& 0 				& 0 		& \ldots 	& 0 \\
e_{1, j_1}^{(0)} 		& e_{1, j_1}^{(1)} 		& 0 		& \ldots 	& \vdots \\
e_{2, j_2}^{(0)} 		& e_{2, j_2}^{(1)} 		& e_{2, j_2}^{(2)} 		& \ddots 	& \vdots \\
\vdots           		& \vdots 			&     		& \ddots	& 0 \\
e_{T-1, j_{T-1}}^{(0)}	& e_{T-1, j_{T-1}}^{(1)}	 &e_{T-1, j_{T-1}}^{(2)}  & \ldots & e_{1, j_{T-1}}^{(T-1)}
\end{bmatrix}, }
\end{align*} \normalsize
where $e_{i, j_i}^{(l)}$ specifies the {($l$+1)}-th digit of event $e_{i, j_i}$, i.e., $d_l$ (cf. Definition \ref{defn:event}).
Using this definition, we impose the following constraint due to delayed data:
\begin{equation}  \hspace{0.07cm}
\mathcal{D}(\mathcal{L}^{E^{\prime}})\hspace{-0.05cm}=\hspace{-0.05cm}\left\{\begin{matrix} \begin{matrix} \{(M^{\alpha}, \\ L^{\alpha})\}^{\vert\mathcal{L}^{E^{\prime}}\vert}_{\alpha=1} \end{matrix} \left \vert 
\begin{matrix}  \forall i,j \in \mathbb{N}_1^{T}:
\\ M^{\alpha}_{ \substack{ ((i-1)n:(i-E_{\alpha}(i,j))n-1, \\ (j-1)p:(j-E_{\alpha}(i,j))p-1)} } \hspace{-0.05cm}=\hspace{-0.05cm} 0, \\
L^{\alpha}_{\substack{ ((i-1)n:(i-E_{\alpha}(i,j))n-1, \\ (j-1)p:(j-E_{\alpha}(i,j))p-1)} }\hspace{-0.05cm}=\hspace{-0.05cm} 0 
\end{matrix}  \right.
\end{matrix}\right\}\hspace{-0.05cm}. \hspace{-0.2cm}
\end{equation} \normalsize

Next, we provide examples of 
$\mathcal{C}(\mathcal{L}^{E^{\prime}})$ and $\mathcal{D}(\mathcal{L}^{E^{\prime}})$. \vspace{-0.1cm}

\begin{example}
Consider two event sequences $\mathcal{E}_1^{\prime} = \{ e_{0,1}e_{1,3} \}$ and $\mathcal{E}_2^{\prime} = \{ e_{0,1}e_{1,2} \}$ of $\mathcal{L}^{E^{\prime}}$ in Example \ref{exm3}. The sets of all non-empty prefixes of $\mathcal{E}_1^{\prime}$ and $\mathcal{E}_2^{\prime}$ are $Pref(\mathcal{E}_1^{\prime}) = \{e_{0,1} , e_{0,1}e_{1,3} \}$ and $Pref(\mathcal{E}_2^{\prime}) = \{e_{0,1} , e_{0,1}e_{1,2} \}$, respectively. It is clear that $\mathcal{E}_1^{\prime}$ and $\mathcal{E}_2^{\prime}$ have the same length 1 prefix, so {we need to impose the following constraints in $\mathcal{C}(\mathcal{L}^{E^{\prime}})$:} $\mathcal{BM}_1(M^1) = \mathcal{BM}_1(M^2) $, $\mathcal{BM}_1(L^1) = \mathcal{BM}_1(L^2) $ and ${\nu}^1_{1:n}= {\nu}^2_{1:n}$. 
Moreover, to formulate the constraints {$\mathcal{D}(\mathcal{L}^{E^{\prime}})$}, 
we first construct 
event matrices associated with $\mathcal{E}^{\prime}_1$ and $\mathcal{E}^{\prime}_2$:
\begin{align*}
E_1 = \begin{bmatrix} e_{0,1}^{(1)} & 0 \\ e_{1,3}^{(1)} & e_{1,3}^{(2)}  \end{bmatrix} = \begin{bmatrix}  1 & 0 \\ 1 & 1 \end{bmatrix}, E_2 = \begin{bmatrix} e_{0,1}^{(1)} & 0 \\ e_{1,2}^{(1)} & e_{1,2}^{(2)}  \end{bmatrix} = \begin{bmatrix}  1 & 0 \\ 1 & 0 \end{bmatrix},
\end{align*}
where a zero element located at the $i$-th row and $j$-th column of matrices $E_1$ (or $E_2$) indicates that the data of time step $j$ is not available at the time $i$ due to the delay in $\mathcal{E}^{\prime}_1$ (or $\mathcal{E}^{\prime}_2$) and causality. This is captured in $\mathcal{D}(\mathcal{L}^{E^{\prime}})$ by $M^{1}_{(0:n-1,p:2p-1)} =0 $ for $i = 1$ and $j =2$ in $E_1$,  $M^{2}_{(0:n-1,p:2p-1)} =0 $ for $i = 1$ and $j =2$ in $E_2$,  and $M^{2}_{(n:2n-1,p:2p-1)} =0 $ for $i = 2$ and $j =2$ in $E_2$, {which results in $M^1$ and $M^2$ with the following block structures:
$M^1=\begin{bmatrix} * & 0 \\ * & *\end{bmatrix}, \ M^2=\begin{bmatrix} * & 0 \\ * & 0\end{bmatrix},$
where $*$ denotes non-zero submatrices.} {Similar constraints also need to be imposed on $L^1$ and $L^2$ in $\mathcal{D}(\mathcal{L}^{E^{\prime}})$.}\vspace{-0.1cm}
\end{example}

%% file: my_theorem.tex
{\subsubsection{Estimator Gains Design}
Next,} we {present an approach to obtain} the estimator gains $(M^{\alpha},L^{\alpha},\nu^{\alpha})$ associated with each unique event sequence $\mathcal{E}_{\alpha}^{\prime}$ in $\mathcal{L}^{E^{\prime}}$ for the estimator in \eqref{eq:estimator}. {Moreover, 
we also allow the time-varying intermediate levels to be prefix-dependent, i.e., with  $\mu_{2,k}^\alpha$, which can lead to improved estimation error bound when the prefix, i.e., the history of available data, is observed at run time. }\vspace{-0.1cm}

\vspace{-0.1cm}
\begin{thm}[Equalized Recovery Estimator Design with Delays] \label{th:delayed} 
	For a system with measurement delays and missing data patterns defined by a fixed-length language $\mathcal{L}$ {given in \eqref{eq:system}, the finite-horizon affine estimator given in \eqref{eq:estimator}} can fulfill the objectives in Problem \ref{prob:EstimatorDesign} if the following is feasible:
	
	\vspace{-0.4cm}\begin{align}
	\label{pb:solve_F_nu}
	\hspace{-0.55cm}\begin{array}{ll}
	\displaystyle\min_{M^{\alpha},\nu^{\alpha},{\mu^\alpha_2},s_0,L^{\alpha}, {\mu_1}}\hspace*{-0.5cm} & \hspace*{0.5cm} {J(\mu_1,\{\mu^\alpha_2\}_{\alpha=1}^{| \mathcal{L}^{E^{\prime} }|})}  \\
	\text{subject to}      
	& \hspace*{-0.6cm}  \forall (\|w\|\hspace*{-0.075cm} \leq\hspace*{-0.075cm} \eta_w,\|v\|\hspace*{-0.075cm}\leq\hspace*{-0.075cm} \eta_v,\|\tilde{x}_{0}\|\hspace*{-0.075cm}\leq\hspace*{-0.075cm} \mu_1, 
	 \alpha \hspace*{-0.075cm}\in\hspace*{-0.075cm} \mathbb{N}_1^{| \mathcal{L}^{E^{\prime} }|}\hspace*{-0.05cm}: \\
	&\hspace*{-0.6cm}  \|\tilde{x}^{\alpha}\| \leq {\mu_2^{\alpha}}, \|R_{T} \tilde{x}^{\alpha}\| \leq \mu_1, {\mu_2^{\alpha}\ge \mu_1, \mu_1 \ge 0,}\\
	&\hspace*{-0.6cm}  \tilde{x}^{\alpha} = \Theta^{\alpha} w + \Psi^{\alpha} v + \Xi^{\alpha} \tilde{x}_0 + \Upsilon^{\alpha} s_0 + {H}\nu^{\alpha}\hspace{-0.05cm},\\
	&\hspace*{-0.6cm}  (M^{\alpha}, L^{\alpha}, \nu^{\alpha}) \in \mathcal{C}(\mathcal{L}^{E^{\prime}}) \wedge \mathcal{D}(\mathcal{L}^{E^{\prime}}),
	\end{array}\hspace{-0.5cm} 
	\end{align}\normalsize 
	
	\noindent  where \vspace*{-0.15cm}
	\begin{align} \label{mp:a_b_r_p}
	\begin{array}{ll}
	R_T &= \begin{bmatrix}0_{n \times nT} & I_n\end{bmatrix}, \  {\mu_2^\alpha}  {= \begin{bmatrix} \mu^\alpha_{2,0},\mu^\alpha_{2,1},\ldots,\mu^\alpha_{2,T}\end{bmatrix}^\top,}\\
	\Theta^{\alpha}   &= (I+{H}(M^{\alpha}+L^{\alpha})C)\Gamma^{\alpha} W,				\\
	\Psi^{\alpha}   &= ({H}(M^{\alpha}+L^{\alpha})(I-C \Gamma^{\alpha} L^{\alpha})- \Gamma^{\alpha} L^{\alpha}) V,	\\
	\Xi^{\alpha} &= (I+{H}(M^{\alpha}+L^{\alpha})C)\Phi^{\alpha},		\				
	{\Upsilon^{\alpha}  = A - \Xi^{\alpha}.}	
	\end{array}
	\end{align}\normalsize
	The matrices {$H$}, $M^{\alpha}$, $L^{\alpha}$, $C$, $\Gamma^{\alpha}$, $W$, $V$, $\Phi^{\alpha}$ and $A$, all of which are stacked matrices for the whole time horizon $T$, are derived after stacking the system in \eqref{eq:system}, estimator in \eqref{eq:estimator} and the output error injection term in \eqref{eq:affine_inp}. 
	The 
	definitions of these matrices can be found in the {Appendix}.\vspace*{-0.1cm}
\end{thm}
\begin{proof}
{The estimator design follows similar steps to the design in \cite{Hassaan2019ACC}. 
It is straightforward to observe that the estimator solves Problem \ref{prob:EstimatorDesign} by construction with} the additional constraints on the estimator gains in Section \ref{sec:EstimatorIntro}. \vspace{-0.1cm}
\end{proof}

{When compared to our prior work \cite{Hassaan2019ACC}, we consider a prefix-based design that enables adaptation of the gain matrices and improved estimation error bounds based on observed prefix, i.e., the history of available data, at run time. It is also noteworthy that in contrast to existing equalized recovery estimators \cite{Rutledge2018,Hassaan2019ACC,Kwesi2019ACC,rutledge2020finite}, the proposed estimator considers $\mu_1$ as a decision variable, instead of a given parameter. This seemingly small change has an important implication that the difficulty in finding the global minimum for $\mu_2$ with previous designs (due to their non-monotonicity in $\mu_1$) can now be overcome with the new design.  }

%% file: my_robustification.tex
{\subsubsection{Robustification}
Next,} since the problem in Theorem \ref{th:delayed} involves semi-infinite constraints (i.e., \emph{for all} constraints), {as in \cite{Hassaan2019ACC}, we leverage robust optimization tools, e.g., \cite{ben2009robust},  
to obtain a problem with a finite number of constraints:}\vspace{-0.1cm} 

\begin{prop}[Robustified Equalized Recovery Estimator Design with Delays]
{The equalized estimator design that solves Problem \ref{prob:EstimatorDesign} via Theorem \ref{th:delayed} is equivalent to:} 

\phantom{a}
\small\vspace{-0.45cm}
\begin{equation} \label{pb:robust}
\begin{array}{l}
\hspace{-0.2cm}\begin{array}{ll} 
\displaystyle\min_{\substack{M^{\alpha},\nu^{\alpha},{\mu_2^\alpha,\mu_1,}s_0,L^{\alpha},\Pi_1^{\alpha},\Pi_2^{\alpha}}} & {J(\mu_1,\{\mu^\alpha_2\}_{\alpha=1}^{| \mathcal{L}^{E^{\prime} }|})}  
\end{array}
\\ \hspace*{-0.2cm}
\begin{array}{l}
\text{subject to}\quad   \Pi_1^{\alpha} \geq 0, \Pi_2^{\alpha} \geq 0,  {\Pi_3^{\alpha} \geq 0, \;} {\mu_2^{\alpha}\ge \mu_1, \; \mu_1 \ge 0,}\\
{\begin{bmatrix}\Pi_1^{\alpha}& \Pi_2^{\alpha} & \Pi_3^{\alpha} \end{bmatrix}}\hspace*{-0.05cm}
\begin{bmatrix}\eta_w \mathds{1} \\ \eta_v \mathds{1} \\ {\mathds{1}} \end{bmatrix} \hspace*{-0.1cm}\leq \hspace*{-0.1cm} \begin{bmatrix} {\mu_2^\alpha} \\ {\mu_2^\alpha}\\ 
\mu_1 \mathds{1} \end{bmatrix} \hspace*{-0.1cm}-\hspace*{-0.1cm}\begin{bmatrix} I & 0\\-I & 0\\ 0 & I \\ 0 & -I\end{bmatrix}  \begin{bmatrix} {H} \nu^{\alpha} \hspace*{-0.075cm}+\hspace*{-0.075cm} \Upsilon^{\alpha} s_0\\ R_{T} ({H} \nu^{\alpha} \hspace*{-0.075cm}+\hspace*{-0.075cm} \Upsilon^{\alpha} s_0)\end{bmatrix}\hspace*{-0.05cm},\\[-3pt]
{\begin{bmatrix}\Pi_1^{\alpha}& \Pi_2^{\alpha} & \Pi_3^{\alpha} \end{bmatrix}}\hspace*{-0.05cm} \begin{bmatrix} I & 0 & 0 \\ -I & 0 & 0 \\ 0 & I & 0\\ 0 & -I & 0 \\ 0 & 0 & I\\ 0 & 0 & -I \end{bmatrix}\hspace*{-0.1cm}=\hspace*{-0.1cm}\begin{bmatrix} I & 0\\-I & 0\\ 0 & I \\ 0 & -I\end{bmatrix}\hspace*{-0.05cm} \begin{bmatrix} G^{\alpha}\\ R_T G^{\alpha}\end{bmatrix} \hspace*{-0.05cm}{\begin{bmatrix} I & 0 & 0\\ 0 & I & 0 \\ 0 & 0 & \mu_1\mathds{1} \end{bmatrix}}\hspace*{-0.05cm},  
\end{array}\\
\; (M^{\alpha}, L^{\alpha}, \nu^{\alpha}) \in \mathcal{C}(\mathcal{L}^{E^{\prime}}) \wedge \mathcal{D}(\mathcal{L}^{E^{\prime}}),\vspace*{-0.4cm}
\end{array}
\end{equation}\normalsize

\vspace*{0.2cm}
\noindent where $G^{\alpha}\triangleq \begin{bmatrix} \Theta^{\alpha} & \Psi^{\alpha} & \Xi^{\alpha}\end{bmatrix}$ with $\Theta^{\alpha}$, $\Psi^{\alpha}$, $\Xi^{\alpha}$, $\forall \alpha \in \mathbb{N}_1^{| \mathcal{L}^{E^{\prime} }|}$ defined in \eqref{mp:a_b_r_p}, while $\Pi_1^{\alpha}$, {$\Pi_2^{\alpha}$,  $\Pi_3^{\alpha}$} are dual matrix variables. 
\end{prop}
{\begin{proof} By replacing the semi-infinite constraints in \eqref{pb:solve_F_nu} with their robust counterparts based on \cite{ben2009robust}, we obtain a similar problem as in \cite[Eq. (11)]{Hassaan2019ACC}. However, since $\mu_1$ is a decision variable in our problem (instead of a parameter as in \cite{Hassaan2019ACC}), we have a bilinear term in the first equality that is a product of  
\syong{dual variables $\tilde{\Pi}_3^\alpha$} with $\mu_1$. To overcome this issue, we post-multiply the second equation on both sides with {\small$\begin{bmatrix} I & 0 & 0\\ 0 & I & 0 \\ 0 & 0 & \mu_1\mathds{1} \end{bmatrix}$\normalsize}, which results in the appearance of the same bilinear term. Then, \syong{since the original $\tilde{\Pi}_3^\alpha$ no longer appears independently, a common trick is to replace $\mu_1\tilde{\Pi}_3^\alpha$} 
with a new decision variable $\Pi^\alpha_3$ that is positive since $\mu_1 \ge 0$. 
\vspace{-0.15cm} \end{proof}

The above optimization problem 
has bilinear terms but is relatively sparse, so off-the-shelf  solvers, e.g., \cite{ipopt}, can find optimal solutions quickly. 
Further, if desired, we can fix $L^\alpha$ and $s_0$  
to perform a line search over $\mu_1$ using 
a linear program without 
loss of optimality, as discussed 
in \cite[Section IV-C]{Hassaan2019ACC}.}

{\subsubsection{Implementation Strategy} \label{sec:imp} 
The} proposed equalized recovery estimator in this paper could be implemented in multiple different ways. 
First, in the case that the delayed/missing data pattern is periodic with a period of $T$ time steps, we can use the same gains for each period because the estimation error bound
at the end of the period is enforced to be the same at the beginning of the period by the proposed estimator.
Moreover, if there is no missing/delayed data, 
we could use an 
equalized performance estimator, i.e., an equalized recovery estimator with a period of 1 time step (cf. \cite{blanchini2012convex}) 
until 
a missing/delayed data is encountered, at which point 
we can switch 
to an equalized recovery estimator with a $T$-length language in which the first data is missing/delayed.
Then, after the recovery time $T$, we revert to the equalized performance estimator again until the next time a delayed data is detected. {Further, if the initial estimation error does not satisfy the equalized recovery/performance level, the proposed estimator can also be combined with any asymptotic estimator, where the latter is used until the desired equalized level is achieved.} 
\syong{Alternatively, we can modify our estimator by replacing $\mu_1$ in the second constraint in \eqref{pb:robust} with the initial estimation error and repeating the process, as needed, to achieve this.}

%% file: my_example.tex
\vspace{-0.05cm}
\section{Examples And {Comparisons}} 
\label{sec:acc_eg} \vspace{-0.05cm}
In this section, the performance of the proposed estimator is validated and compared with {the approaches in \cite{zhang2011linear} and \cite{Kwesi2019ACC}}. 
The examples using our proposed estimator are all run using MATLAB 2017a. 
As the robustified problem in \eqref{pb:robust} involves many sparse matrices, the IPOPT solver \cite{ipopt} is used. Moreover, in \cite{Hassaan2019ACC}, it was {established}
that the value of $s_0$ \hasnb{in \eqref{pb:robust}} does not affect the performance of the estimator. So, to simplify the problem of the estimator design, all of the parameters of $s_0$ are set to zero in all the presented examples. \vspace*{-0.45cm}

\input{my_brp}

\input{my_acc}

%% file: my_brp.tex
\vspace{-0.1cm}\subsection{Batch Reactor Process 
{(Comparison with \cite{zhang2011linear})}}\label{sec:brp_eg} \vspace{-0.1cm}
To demonstrate the capability of the proposed estimator proposed in this paper { in comparison with 
\cite{zhang2011linear}} {when output delays are involved}, we utilize an example of a continuous-time batch reactor process 
from \cite{shoukry2013minimax}. 
This system is first discretized with a sampling time of {$T_s$ = 0.05} seconds using MATLAB {\texttt{c2d}} command \hasnb{\syong{(with zero-order hold)}
} {to obtain a discrete-time state-space system with the following matrices:} 

{
\small 
\vspace{-0.35cm}
\begin{gather*} \label{model: BRP}
\hspace{-0.1cm}\begin{array}{c} 
A \hspace*{-0.1cm}=\hspace*{-0.175cm} \left[\setlength\arraycolsep{0.3pt}
\begin{array}{cccc}
    1.0795 &  -0.0045 &   0.2896 &  -0.2367 \\
   -0.0272 &   0.8101 &  -0.0032 &   0.0323 \\
    0.0447 &   0.1886 &   0.7317 &   0.2354 \\
    0.0010 &   0.1888 &   0.0545 &   0.9115
\end{array}
\right]\hspace*{-0.125cm}, B \hspace{-0.1cm}=\hspace{-0.175cm} \left[\setlength\arraycolsep{0.7pt}
\begin{array}{cc}
    0.0006 &  -0.0239 \\
    0.2567 &   0.0002 \\
    0.0837 &  -0.1346 \\
    0.0837 &  -0.0046
\end{array}
\right]\hspace{-0.125cm},
\end{array}\\[-0.1cm]
C \hspace{-0.05cm}=\hspace{-0.05cm} \left[
\begin{array}{cccc}
    1 & 0 & 1 & -1 \\
	0 & 1 & 0 &  0
\end{array}
\right]\hspace{-0.05cm}, 
V = I_p,
W = \emptyset.
\end{gather*}\vspace*{-0.4cm}  \normalsize
 }

The time horizon is taken to be $T = 5$ with maximum possible delay of 2 steps.
This results in the delayed data model that can be expressed as the fixed-length language containing $3^5$ words, i.e. $\mathcal{L} = \{\mathcal{W}_1, \ldots, \mathcal{W}_{243}\}$. For the proposed estimator, according to Definitions \ref{defn:event}--\ref{defn:eventseq}, we can find the corresponding event-based language $\mathcal{L}^{E}$ as well as the reduced language $\mathcal{L}^{E^{\prime}}$.
{The measurement noise bound $\eta_v=0.05$ is chosen to cover 5 standard deviations of $v \sim \mathcal{N}(0,0.01^2)$, {and by} 
solving the robustified problem \eqref{pb:robust} {with the cost function $J(\cdot)=\mu_1+\sum_{k=0}^T \sum_{\alpha=1}^{| \mathcal{L}^{E^{\prime} }|} \mu^\alpha_{2,k}$}, we obtain 
{recovery levels of} $\mu_1 = 0.33$ and $\max_{k,\alpha}(\mu_{2,k}^\alpha) = 0.6912$.}

\begin{figure}[!t]
	\centering
	\begin{subfigure}[t]{0.24\textwidth}
		\centering
		\includegraphics[scale=0.445,trim=0mm 3mm 0mm 0mm]{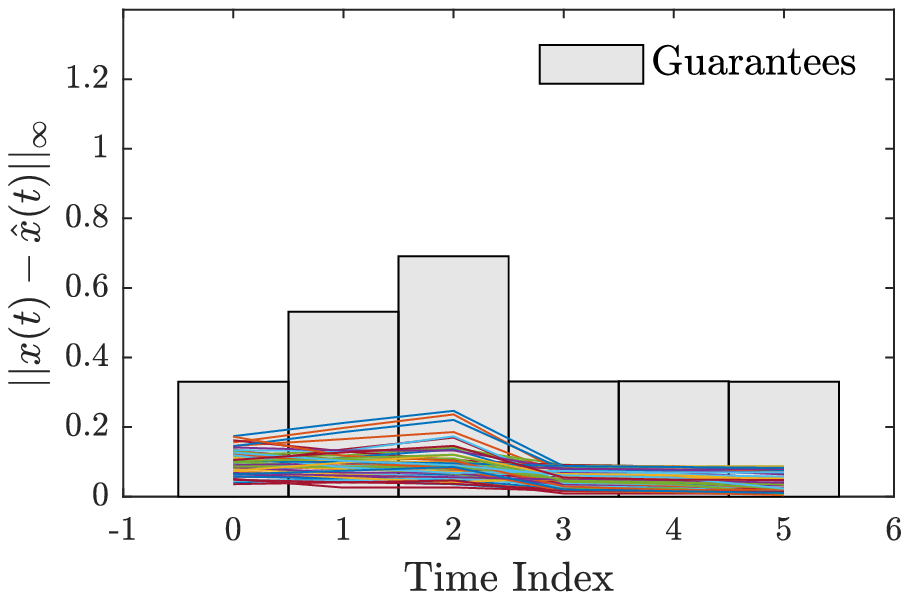}
		\caption{{Proposed Estimator.}}
		\label{subfig:sae_prop}
	\end{subfigure}%
	~
	\begin{subfigure}[t]{0.24\textwidth}
		\centering
		\includegraphics[scale=0.445,trim=0mm 3mm 0mm 0mm]{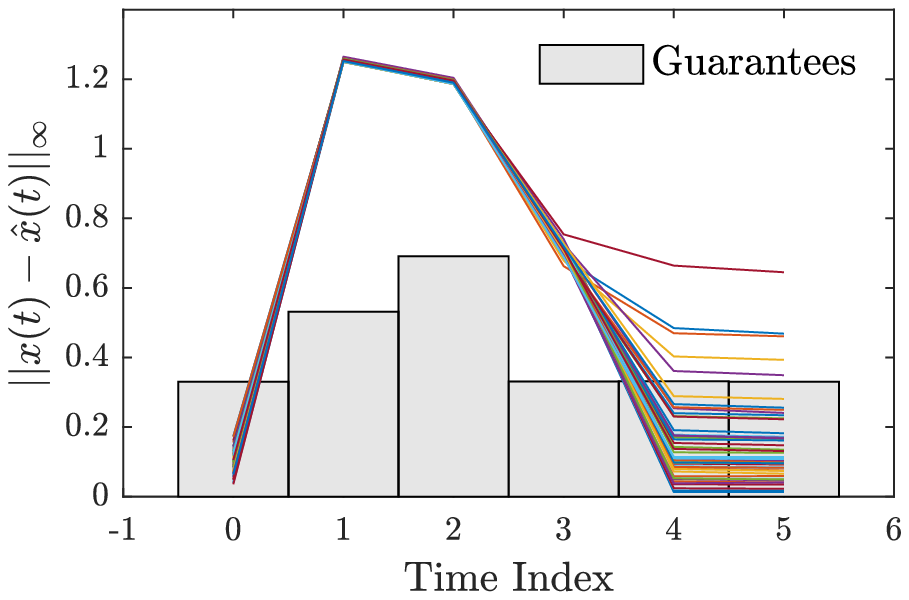}
		\caption{{Estimator from \cite{zhang2011linear}.}}
	\end{subfigure} \vspace{-0.15cm}
	\caption{{Estimator comparison for {$\mathcal{W}_{sim} = 21210$}.}\label{fig:sae_eg}} \vspace{-0.235cm}
\end{figure}

{To compare the performance of our proposed design with the Kalman filter based estimator design for systems with delayed data in \cite{zhang2011linear}, we let {$x(0) = [1,1,1,1]^\top$} and the true delay pattern be {$\mathcal{W}_{sim}=21210$}. 
{For the simulation,} we randomly generated the initial state error and noise signals from truncated normal distributions with zero means and covariance matrices {$P_0=(\mu_1/5)^2 I_4$, $Q=\emptyset$} and {$R=(\eta_v/5)^2  I_2$, where the initial error and noise bounds, $\mu_1$, $\eta_v$, represent 5 times their standard deviations.} 
{Figure \ref{fig:sae_eg} shows the results of 50 runs, where the estimation errors using the proposed estimator stay within the guaranteed bounds, as desired, and are much less than the estimation errors 
from \cite{zhang2011linear}, which are not 
within the bounds, {as one may expect.}}}

%% file: my_acc.tex
\vspace{-0.15cm}
\subsection{Adaptive Cruise Control 
(Comparison with \cite{Kwesi2019ACC})} 
\begin{figure}[!t]
	\centering 
	\begin{subfigure}[t]{0.24\textwidth}
		\centering
		\includegraphics[scale=0.445,trim=0mm 3mm 1mm 0mm]{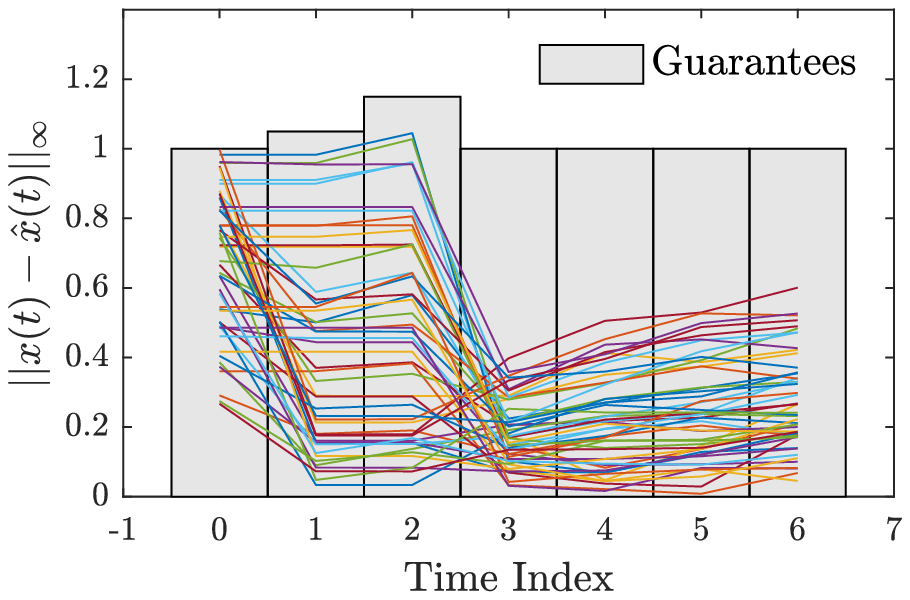}
		\caption{Proposed estimator.}
	\end{subfigure}%
	~
	\begin{subfigure}[t]{0.24\textwidth}
		\centering
		\includegraphics[scale=0.445,trim=0mm 3mm 0mm 0mm]{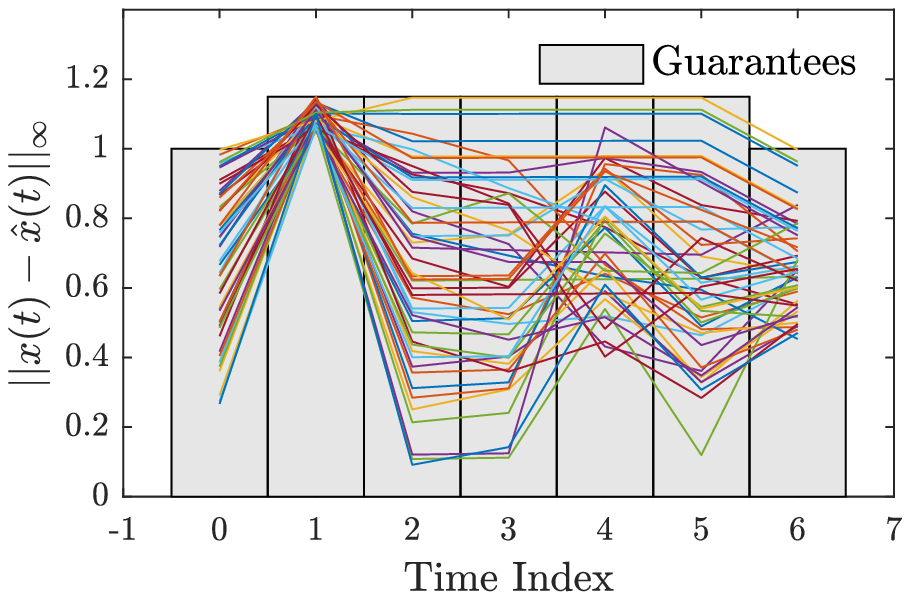}
		\caption{Estimator in \cite{Kwesi2019ACC}.}
	\end{subfigure} \vspace{-0.15cm}
	\caption{Estimator comparison for missing data at $k=1$.\label{fig:acc_eg}} \vspace{-0.235cm}
\end{figure}

\vspace{-0.1cm}
In the previous example, we showcased the capability of the proposed observer in the case of delay scenarios. 
Since missing data is basically a special case of delays beyond the finite time horizon, we also compare the proposed observer for the missing data scenario with another {missing data (only)} estimator in 
\cite{Kwesi2019ACC}. We use the same example presented in \cite{Kwesi2019ACC} of an adaptive cruise control, with the time horizon of $T=6$ and {for the sake of comparison, the equalized recovery level} 
$\mu_1=1$ is {specified instead of {letting} it be a {decision} variable}\footnote{{This leads to a conservative design since it must satisfy $\mu^\alpha_{2,k} \geq \mu_1=1$. In fact, we obtain  $\max_{k,\alpha}{\mu_{2,k}^\alpha}=0.3991$ without fixing $\mu_1=1$.}}. 
The equivalent language to the one 
in \cite{Kwesi2019ACC} is {used, i.e.,} 
$\mathcal{L} = \{060000,006000,000600,000060\}$. After running the optimization problem, {the maximum value of the intermediate upper bound obtained is $\syong{\mu_2 \triangleq \max_{k,\alpha}{\mu_{2,k}^\alpha}}= 1.1498$, which is the same 
value obtained in \cite{Kwesi2019ACC}. Contrary to \cite{Kwesi2019ACC}, the value of $\mu_2$, being {time-varying} in our approach, is not {always at its} maximum, 
hence guaranteeing less error even during the intermediate phase}. In our simulation, {the true missing data pattern is 
$\mathcal{W}_{sim} = \{060000\}$} that 
corresponds to missing data at $k=1$, and we compared the result with those in \cite{Kwesi2019ACC}. 
A comparison of both the estimators is depicted in Figure \ref{fig:acc_eg} that shows that the proposed estimator 
performs 
{better than 
the} estimator in \cite{Kwesi2019ACC} 
for the missing data scenario.  

